\documentclass[12pt,preprint]{aastex}

\shorttitle{winds of V592 Cas}
\shortauthors{Kafka et al.}
\begin{document}
\title{Observations of V592 Cas - An Outflow at Optical Wavelengths}

\author{Kafka, S.\altaffilmark{1}, Hoard, D.W.}
\affil{Spitzer Science Center/Caltech, 220-6, 1200 E.California Blvd, Pasadena, CA 91125, USA}

\author{Honeycutt, R.K. \and Deliyannis, C.P.}
\affil{Indiana University, Astronomy Department, Swain Hall West, Bloomington, IN 47405, USA}

\altaffiltext{1}{email: stella@ipac.caltech.edu}

\begin{abstract}

We present new red optical spectra of V592 Cas aimed at exploring the properties of the outflow of this system
in a spectral region where the underlying white dwarf and the accretion disk do not contribute significantly to the observed absorption components of the H$\alpha$ and HeI line profiles. We use the H$\alpha$ emission line to study the wind, which appears as pronounced blueshifted P-Cygni absorption troughs whose low velocity end contaminates the blue side of the emission line profile. The wind appears to be episodic in nature, with multiple events reaching velocities of 5000km/sec in H$\alpha$. Similar (but weaker) wind signatures appear in the HeI5876$\AA$ line but are absent in HeI6678$\AA$. Our data suggest that during wind episodes the wind is phase dependent and is visible for half the orbit of the system. Considering that V592 Cas is viewed almost face-on, the symmetry axis of the outflow can not be orthogonal to the disk and/or the outflow must have some other inherent asymmetry in outflow geometry. A possible origin of the wind is in a disk hotspot, either at the initial impact point of the accretion stream on the disk edge or as a result of disk overflow (similar to SW Sextantis stars). Simultaneous optical photometry during one night of our spectroscopic observations indicate that there is no clear relationship between the optical brightness variations and the strength of the outflow in this system.

\end{abstract}

\keywords{stars: individual (V592 Cas), stars: mass loss, (stars:) novae, cataclysmic variables, stars: winds, outflows}

\section{Introduction}

Cataclysmic variables (CVs) are semi-detached binary stars in which a white dwarf (WD) is accreting H or He-rich material from its low mass, usually main sequence companion. In most cases, matter is transferred onto the WD via a accretion disk (disk CVs) whereas when the magnetic field of the WD is strong (B$\ge$10$^{7}$G) an accretion column leads material directly onto the magnetic pole(s) of the WD via its magnetic field lines (magnetic CVs). 

The evolution of CVs and all interacting binary stars (and hence knowledge of their progenitors and descendants) is driven by transfer and loss of mass and angular momentum. According to the standard scenario for CV evolution angular momentum is lost from the system via magnetic braking and/or gravitational radiation (Spruit $\&$ Ritter 1983). Existing models suggest that magnetic braking alone may not provide enough angular momentum loss to account for high mass transfer rate ($\dot{M}$) CVs above the gap (Taam \& Spruit 2001). An alternative mechanism for CV angular momentum loss could be provided by gravitational torques from a circumbinary (CB) disk (Taam \& Spruit 2001).  This mechanism is provided via a disk wind resulting from the accretion process, whose strength depends explicitly on mass transfer (consequential mass loss) and may contribute significantly to the evolution of the system (Cannizzo \& Pudritz 1988). 

 Outflows in CVs have been observed in the UV by the International Ultraviolet Explorer (IUE) satellite and the Space Telescope Imaging Spectrograph (STIS) on the Hubble Space Telescope (HST), as P-Cygni profiles in species like CIV (1548$\AA$ \& 1551$\AA$), NV~(1239$\AA$ \& 1243$\AA$) and SiIV~(1397$\AA$), with maximum observed blueshifted absorption reaching 5000~km~sec$^{-1}$ (e.g. Prinja et al. 2004) indicating wind mass loss rates as high as a percentage of the mass transfer rate ($\dot{M}$$\sim$5$\times$10$^{-10}$M$_{sun}$/yr; eg. Proga 1999). Until recently, only BZ Cam has been known to have reliable P~Cygni profiles at optical wavelengths (Ringwald \& Naylor 1998) as well as in the far-UV (Hollis et al. 1992). However, BZ Cam is an unusual system in many respects (e.g. it is the only CV embedded in a nebula that is not associated with a recorded nova explosion) and may not be representative of disk CVs. Recent investigations have demonstrated that outflows in CVs can be observed in the optical regime using appropriate wind diagnostics. In a spectroscopic study of the old nova Q~Cyg, Kafka et al. (2003) examined time-resolved optical spectra in which pronounced P~Cygni profiles were present in the He I triplet lines at 5876$\AA$ and 7065$\AA$ and in H$\alpha$, indicating the presence of an outflow with velocity reaching 1500 km sec$^{-1}$. Using those lines as wind diagnostics, more disk CVs were discovered to display wind signatures in their optical spectra (Kafka $\&$ Honeycutt 2004). The use of the HeI triplets as outflow tracers provides a new tool for studying CV disk winds and their characteristics using ground-based facilities. Therefore, we initiated a survey of outflows from high mass transfer rate disk CVs, looking for wind signatures in the HeI triplets and/or in the H$\alpha$ line. The primary goal of the survey is to determine if CV winds can provide an effective mechanism of angular momentum loss that contributes to the secular evolution of those systems. 

One target of our survey is V592~Cas, a disk CV discovered by Greenstein, Sargent $\&$ Haug (1970) via its very blue color and spectral characteristics. Spectroscopic observations of this system were among a collection of previously ``poorly studied'' CVs presented in Downes et al. (1995). The spectrum displays strong HeII~4686$\AA$ and Bowen blend (CIII/NIII) emission and weak Balmer and HeI emission lines. The first time-resolved optical spectroscopic study of the system (Huber et al. 1998) derived an initial orbital period of 2.472h. This was later revised by Taylor et al. (1998; hereafter T98) and Witherick et al. (2003; hereafter W03) to P$_{orb}$=2.76h, placing the system toward the long period end of the CV period gap. Although it is not eclipsing (i$\sim$28$\arcdeg$; Huber et al. 1998) interest in V592 Cas increased due to its diverse periodicities; it has both positive and negative superhumps (with periods of P$_{orb}$+0.062d and P$_{orb}$-0.028d; T98) as well as a suggested 21-min photometric oscillation (Kato $\&$ Starkey 2002). In the UV, the system displays a rich suite of emission lines (CIII, PV, SIV OVI and Ly$\beta$) most of which are affected by strong and variable P~Cygni profiles, indicating the presence of a wind reaching velocities of 2500-3000 km/sec (Prinja et al. 2004). An outflow was also found to be present in the optical spectra of the system (T98, W03).

In this paper we present new multi-epoch time-resolved red spectroscopy of V592 Cas in which the underlying absorption features from the white dwarf and the accretion disk are not present. The purpose of our study is to characterize the outflow of the system as it appears in the H$\alpha$ line and the HeI triplets, and examine correlations with variations in the optical light curve. Our observations and data reduction techniques are described in section 2, and our analysis of the data follows in section 3. We finish by reviewing our main conclusions of this work in section 4.

\section{Observations}

\subsection{Optical Spectroscopy}

Our spectroscopic data were obtained using the WIYN\footnote{The WIYN Observatory is a joint facility of the University of Wisconsin-Madison, Indiana University, Yale University, and the National Optical Astronomy Observatory.} 3.5-m telescope/Hydra multi-object spectrograph (MOS) and the KPNO~2.1-m telescope/GoldCam spectrometer during four nights in Sept-Oct~2005. For the Hydra/MOS data, the 600~line~mm$^{-1}$ grating was used in first order, blazed at 7500$\AA$; the spectral coverage was 5500-8000$\AA$, with a resolution of $\sim$3$\AA$. The weather was fair, therefore our exposure times were fixed at 600s. For sky subtraction we used information from numerous unparked fibers; a CuAr lamp was used for wavelength calibration. For the GoldCam observations, we used grating 35 with a spectral coverage of 5300-8600$\AA$ and a resolution of $\sim$3$\AA$. Exposure times ranged between 180 and 300 sec; the shorter exposure times were used in order to resolve rapid changes in the H$\alpha$ profile when the weather permitted. A HeNeAr lamp was employed for wavelength calibration. 

Dome flat fields were obtained for all data sets; flat-field corrections, as well as bias and sky subtraction, were performed using the standard IRAF\footnote{IRAF is distributed by the National Optical Astronomy Observatory, which is operated by the Association of Universities for Research in Astronomy, Inc., under cooperative agreement with the National Science Foundation.} procedures. For the GoldCam data the IRAF packages {\it apextract} and {\it apall} were used for extracting one-dimensional spectra. Reductions of the Hydra/MOS data used the IRAF~{\it twodspec} and {\it onedspec} packages. The reduction steps included extraction of one-dimensional star and sky fiber spectra, flat-field correction, wavelength calibration, correction for scattered light, and sky subtraction. No corrections were made for strong telluric absorption features such as the  a and B bands of O$_{2}$.  However, the relatively weak telluric H$_{2}$O absorption in the region 5900-6020\AA was removed because of its occasional effect on the profile of H$\alpha$.  We interactively corrected the optical spectra using data from the Wallace $\&$ Livingston (2003) atlas and the IRAF task {\it telluric}; a demonstration of the procedure is presented in Kafka, Anderson $\&$ Honeycutt (2008) and is not repeated here. 

\subsection{Optical Photometry}

Simultaneous optical photometry was obtained during the night of 2005 Oct 08 UT using a Johnson V~filter on the S2KB~CCD camera on the WIYN~0.91 m telescope located at Kitt Peak, Arizona. We used 30s for all our exposures. Bias subtraction and dome flat fielding were applied using standard IRAF routines followed by APPhot aperture photometry. Incomplete ensemble photometry (Honeycutt 1992) was then used to reduce the data. The instrumental magnitudes from the WIYN~0.91-m V band photometry were placed on a standard magnitude scale using secondary standards $\#$4 and $\#$14 from Henden $\&$ Honeycutt (1995). Information on this photometry is also included in Table~1.

\section{Analysis}

\begin{figure}
\includegraphics[angle=-90,scale=.70]{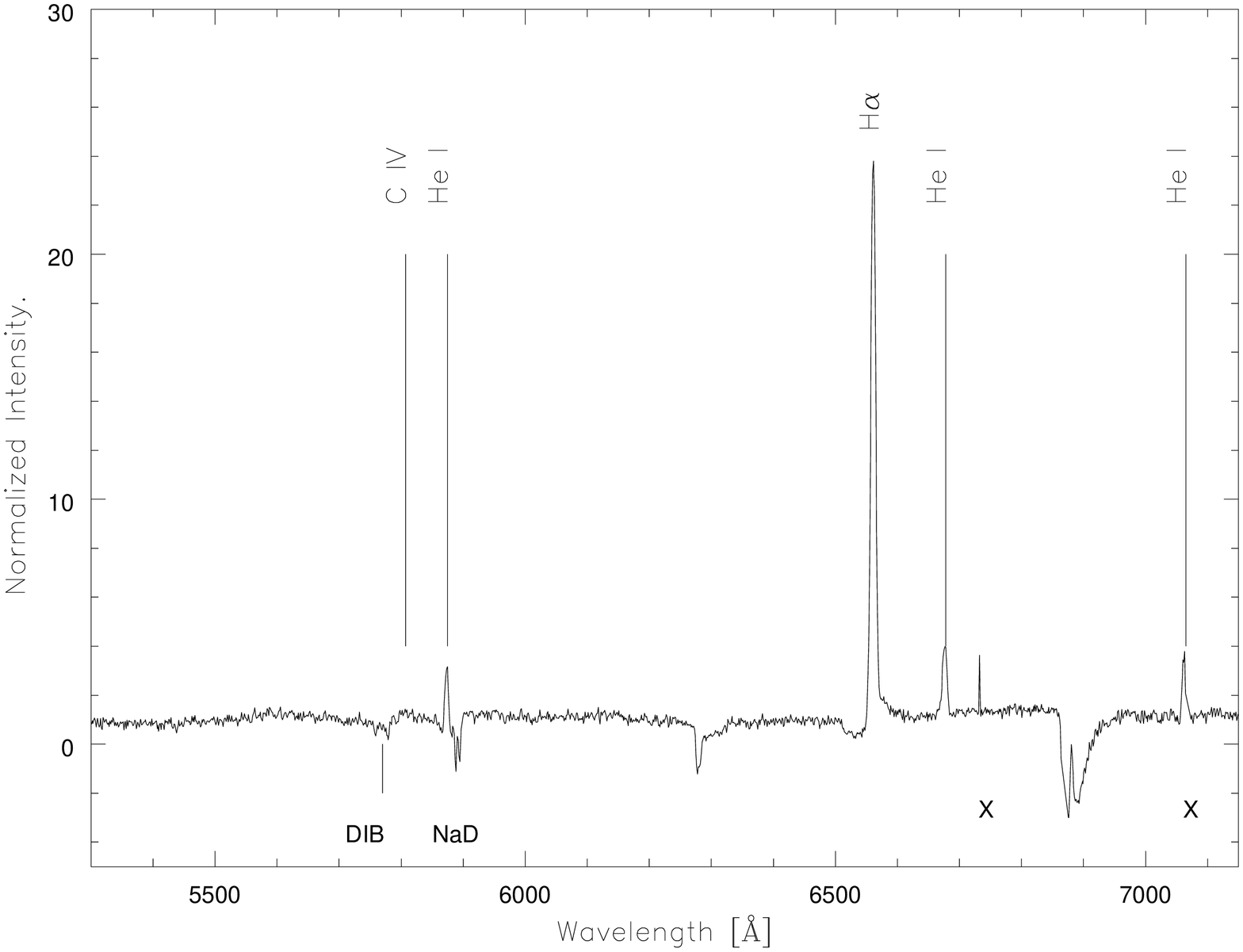}
\caption{Average red spectrum of V592 Cas
with emission/absorption features marked. Defects (blocked columns) in the spectrum that were corrected for presentation purposes are marked with an ``X''. The spectrum is continuum normalized. \label{allspec}}
\end{figure}

Figure~\ref{allspec} presents the average red spectrum of V592~Cas, with the main absorption and emission features labeled; the spectrum is continuum normalized. The equivalent widths (EWs) and full width at half maximum of the Gaussian fits to the major emission features in our spectra are presented in Table~2. 

 V592~Cas resembles a typical low inclination disk-accreting nova-like CV, exhibiting a single-peaked H$\alpha$ emission line and HeI~5876$\AA$, 6678$\AA$ and 7065$\AA$ emission. Noteworthy is the likely detection of CIV (5801,5812$\AA$) emission in our average spectrum. The presence of this line was also reported in T98 and it could indicate the presence of a strong ionization field in the system\footnote{This CIV line likely originates from a CV$\rightarrow$CIV recombination cascade requiring 490eV.} or a C-rich donor star. In our data the detection appears marginal and overlaps with a DIB in the blue end and with the wings of the HeI wind in the red, therefore it is not possible to measure the EW of the line. Also, we don't see any spectral features of the donor star, even when we combine all our individual spectra to maximize the S/N. A striking feature of this red spectrum of V592 Cas is the presence of a deep  blueshifted absorption trough (P-Cygni profile) accompanying the H$\alpha$ emission line; the sharp drop in the blue wing of the emission attests to the wind contaminating the emission line profile at low velocities. The absence of WD/accretion disk absorption contaminating this line makes it an ideal tool for the study of the outflow in V592 Cas. The P-Cygni profiles in the HeI 5876$\AA$ line also help determine characteristics of the wind; however the absorption trough in this line is much weaker than in H$\alpha$. Therefore, H$\alpha$ will be the primary tool for our analysis of the wind characteristics.

 Using the IRAF ``k'' routine in the {\it splot} task and a single Gaussian fit, we measured the radial velocities and equivalent widths of the emission lines of HeI and H$\alpha$. By an unfortunate coincidence, a blocked column on the GoldCam CCD (KPNO 2.1m) intercepted the HeI 7065$\AA$ line. In order to include this line in our analysis we corrected the defect by visual interpolation using IRAF's {\it splot}; nonetheless determination of EW and radial velocities (RV) from this line have larger errors and will not be weighted heavily in our conclusions. To exclude possible radial velocity variations resulting from the wavelength calibration process of our data (due to an instrumental drift on the spectrograph) we measured the radial velocities of the telluric O$_{2}$ (6280$\AA$ and 6867.2$\AA$) bands in the spectra and examined for any variations in the wavelength, finding none of significance.

W03 derived an orbital period for the system based on the H$\beta$, H$\gamma$ and HeII4686$\AA$ emission lines. This HeII line can have a wind origin or be strongly affected by it (Honeycutt et al. 1986; Marsh \& Horne 1990). T98 notice that the Balmer lines are highly variable in their spectra on different nights of their observations due to the presence of a wind, therefore they were not suited for a period study of the system; they used the HeI 5876$\AA$ and 6678$\AA$ emission lines instead. We have used the spectroscopic ephemeris of T98 for our analysis. Efforts to improve the ephemeris of V592 Cas were thwarted by the relatively long intervals between epochs in the literature, leading to cycle count ambiguities using the period errors provided in the literature.  Also, the relatively short (4 week) interval of our data made it impossible to improve the period using our data alone.  We can however provide an updated JD0 = 2,453,654.88$\pm$0.02 valid for the epoch of our data.  This is for the - to + crossing of $\gamma$ for the He I lines, and was calculated by noting that this crossing occurs at phase 0.19$\pm$0.02 (see Fig 2 and Table 3) for our data, using the T98 ephemeris. Because our interest is more in the character of the phase dependences than in the absolute phases, the remaining uncertainties in the ephermeris do not affect our analysis.

\begin{figure}
\epsscale{1.2}
\plottwo{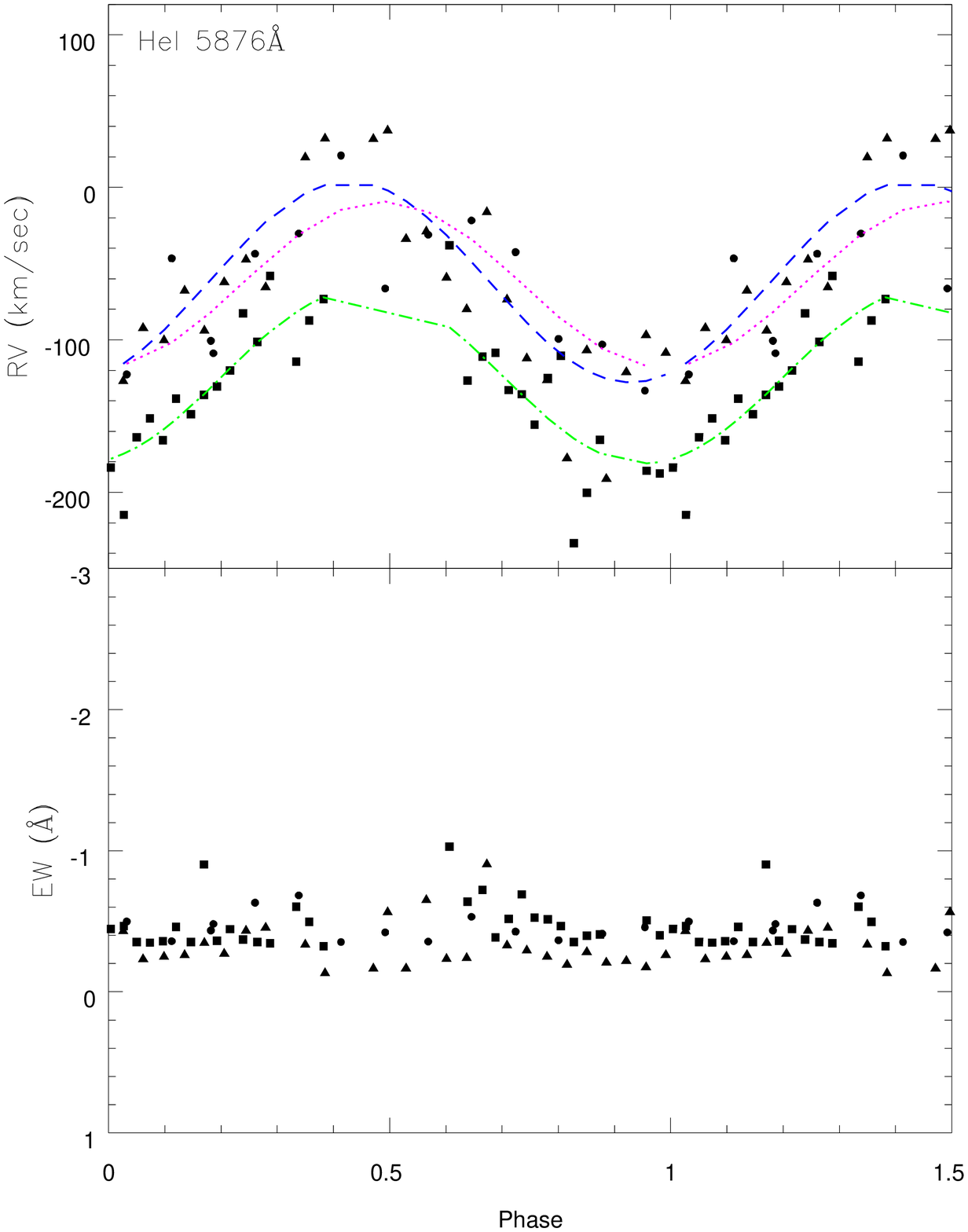}{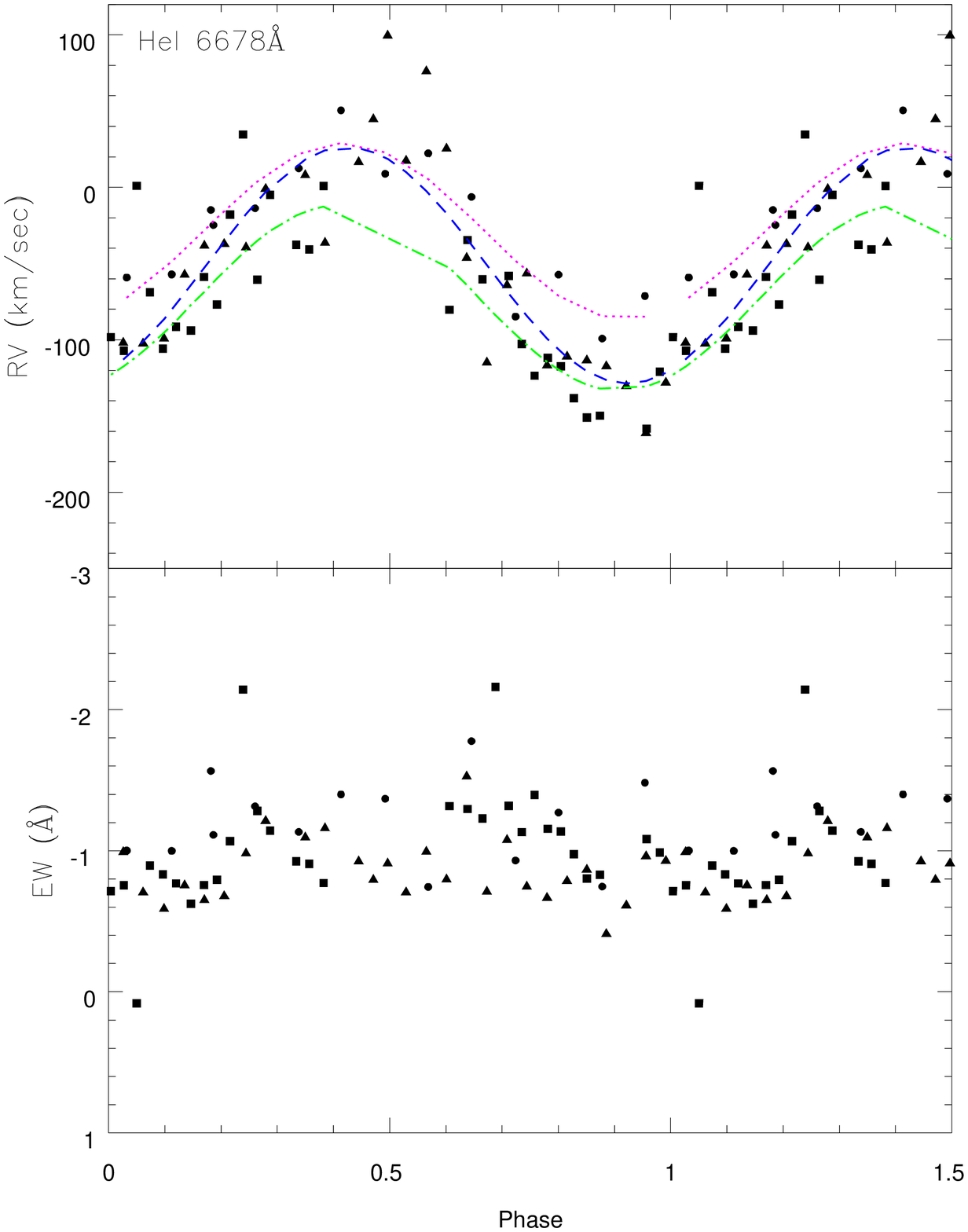}
\caption{Radial velocity (top) and EW (bottom) of HeI 5876$\AA$ and 6675$\AA$ emission lines. Triangles are used for the 2005 Sept 09 data, squares for the 2005 Oct 08 data, and circles for the 2005 Oct 11 data. The parameters of the radial velocity fits at each epoch of our observations are presented in Table 3: the magenta dotted line is for the 2005 Sept 09 data, the blue dashed line for the 2005 Oct 08 data, and the green dot-dashed line for the 2005 Oct 11 data. Note that the assymetry in the 2005 Oct 11 line is due to lack of data between phases 0.4-0.6 on that night of observations. See text for details.\label{HeI}}
\end{figure}

\begin{figure}
\epsscale{.80}
\plotone{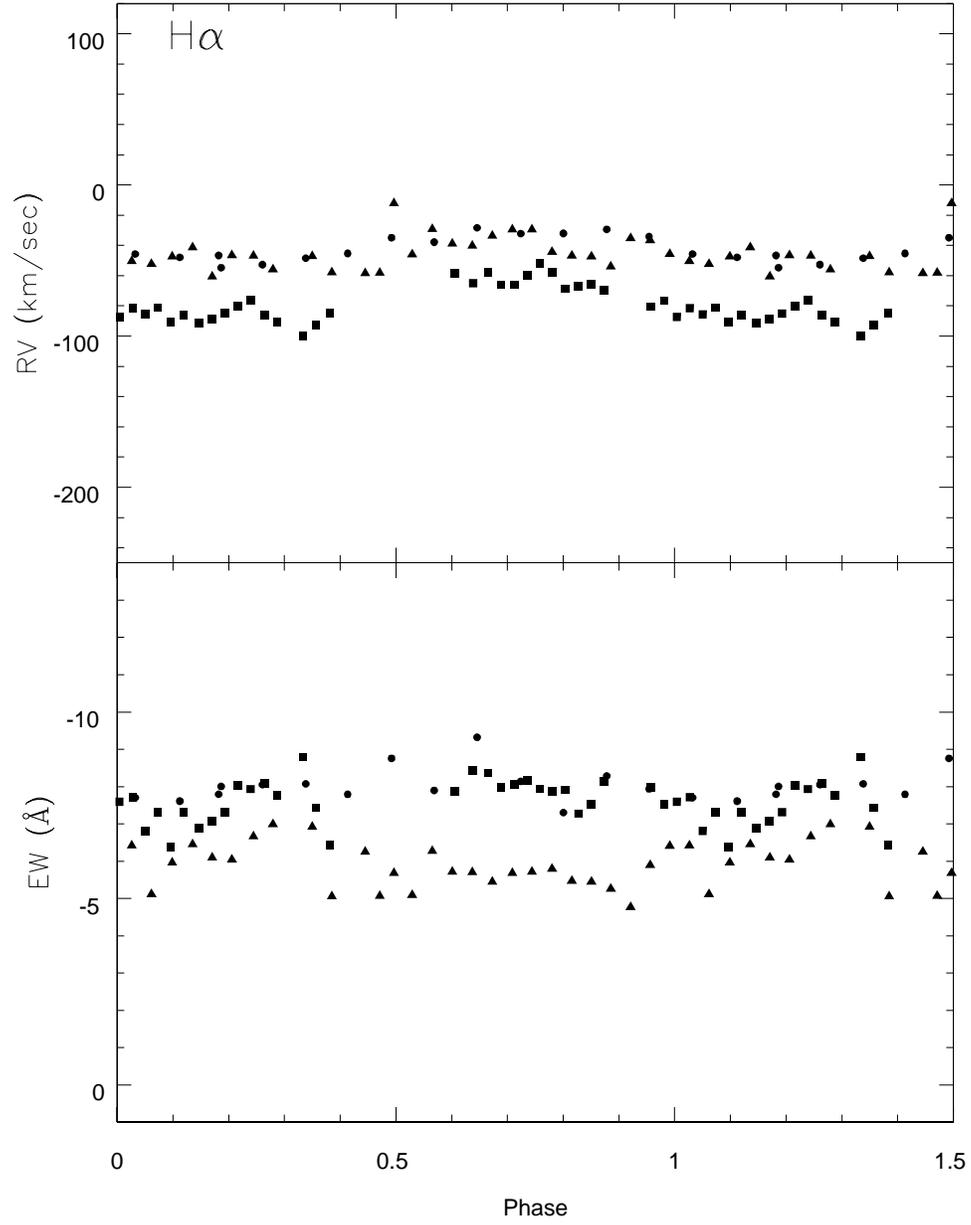}
\caption{Radial velocity and EW of the H$\alpha$ line. The symbols are the same as in figure \ref{HeI}. \label{fullHa}}
\end{figure}

The resulting plots for our RVs of HeI 5876$\AA$, HeI 6678$\AA$ and H$\alpha$ emission lines appear in Figures~\ref{HeI} and \ref{fullHa} (the HeI 7065$\AA$ line exhibits a similar behavior to the other HeI lines, albeit with more scatter). All HeI lines appear to be sinusoidally modulated with the orbital phase. Using sinusoids of the form {\it v(t) = $\gamma$ + Ksin[2$\pi$($\phi$ - $\phi_{0}$)]} in individual radial velocity curves from each night we derived the systemic velocity ($\gamma$), the semi-amplitude of the sinusoidal fit (K) and the gamma-crossing phase ($\phi_{0}$) values listed in table~3. Note the changes in $\gamma$ of the HeI emission lines on different nights (Table 3).  Such changes are often seen in NL CVs (especially in SW Sex stars), and may be due to a varying wind contributions to the line profiles.

\begin{figure}
\epsscale{1.0}
\plotone{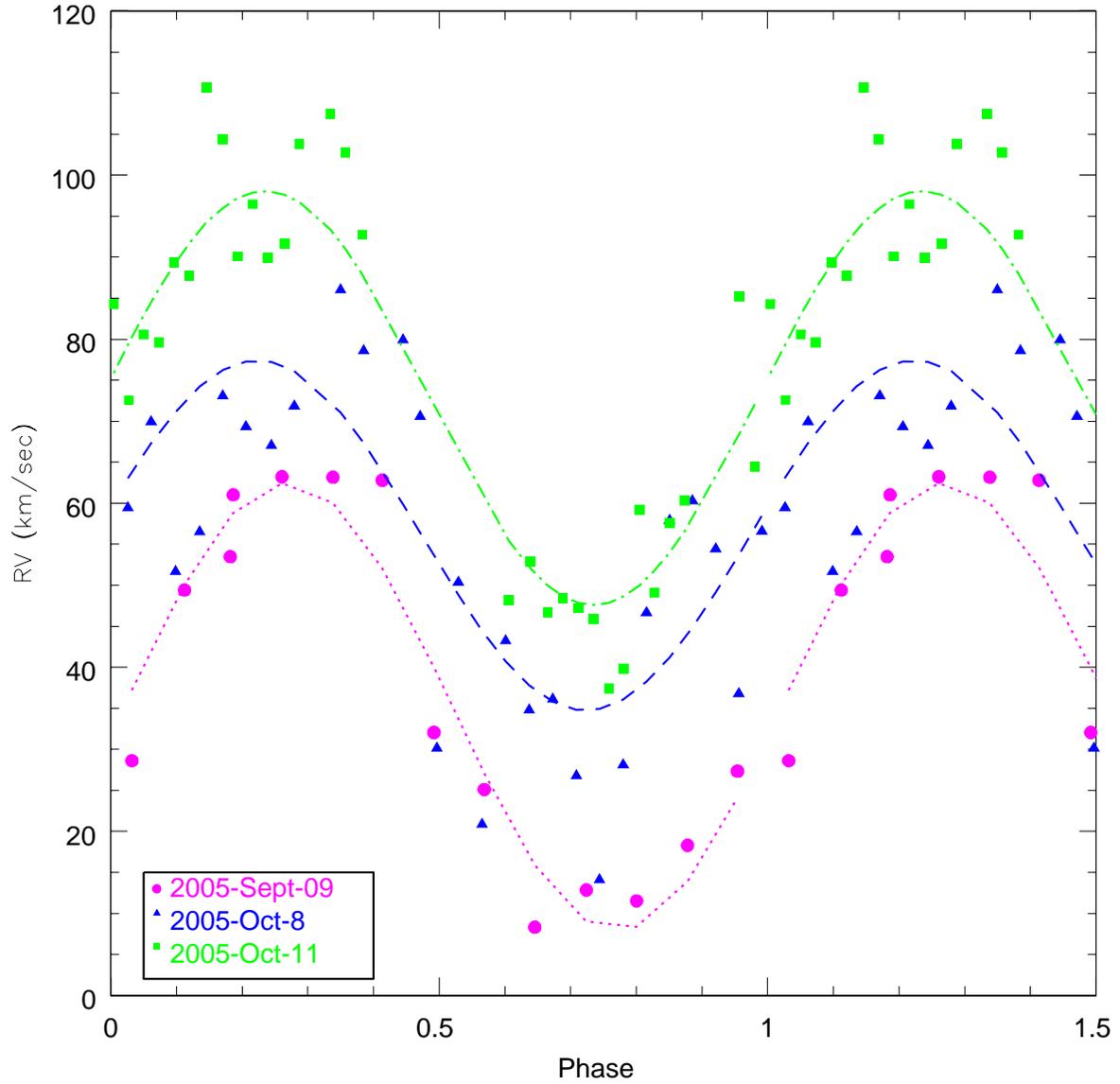}
\caption{RV and fit of the central peak of H$\alpha$ for three epochs of observations. Although all fits are in phase (indicating common origin of the central emission), the $\gamma$ and K velocity vary. This demonstrates the presence of a variable component resulting in a variable central peak of H$\alpha$ at different epochs.\label{centerHa}}
\end{figure}

The low inclination angle of V592 Cas allows visibility of all emission sites (accretion disk, hotspot, white dwarf, irradiated donor star) at all phases. For each spectral feature we measured, the K velocities are similar (within error) at all epochs, as is the phasing of the orbital modulation; this allows grouping data of all nights to derive properties of the system. Note that there is no detectable orbital modulation of the H$\alpha$ emission line (fig~\ref{fullHa}). This comes as a surprise since the RV of the Balmer emission lines (H$\beta$ and H$\gamma$) in V592~Cas show strong sinusoidal modulation with velocities reaching K$\sim$50-km/sec (W03).
In Figure~\ref{fullHa} it appears that the systemic velocity of the H$\alpha$ emission line changed by $\sim$40 km s$^{-1}$ between nights.  Based on the stability of the RV of telluric features, we think this detection is real. Because the winds in CVs are often highly variable, it is very likely that the apparent changing gamma is due to a varying wind contribution that affects the shape (and enhances the apparent gamma velocity) of the line profile from night-to-night. Focusing on the top 30$\%$ of the H$\alpha$ line (fig~\ref{centerHa}) reveals an s-wave shifted by $\sim$0.1 phase units with respect to the HeI lines, with a variable systemic velocity for each of the nights of our observations (see Table~3). A phase shift between the RVs of the Balmer and HeI lines is a characteristic commonly present in SW~Sex type nova-like CVs (e.g. V442~Oph; Hoard, Thorstensen \& Szkody 2000) making V592 Cas a potential member of that class of CVs (although both T98 and W03 argue against this possibility and we acknowledge the difficulty of confirming a low-inclination CV a a member of a group whose observational characteristics are most apparent in higher inclination CVs). Interestingly, W03 demonstrate that their H$\beta$, HeI4471 and HeII4686 also show significant levels of variability near the line centers by about $\sim$2000km/sec.

\begin{figure}
\epsscale{1.3}
\plottwo{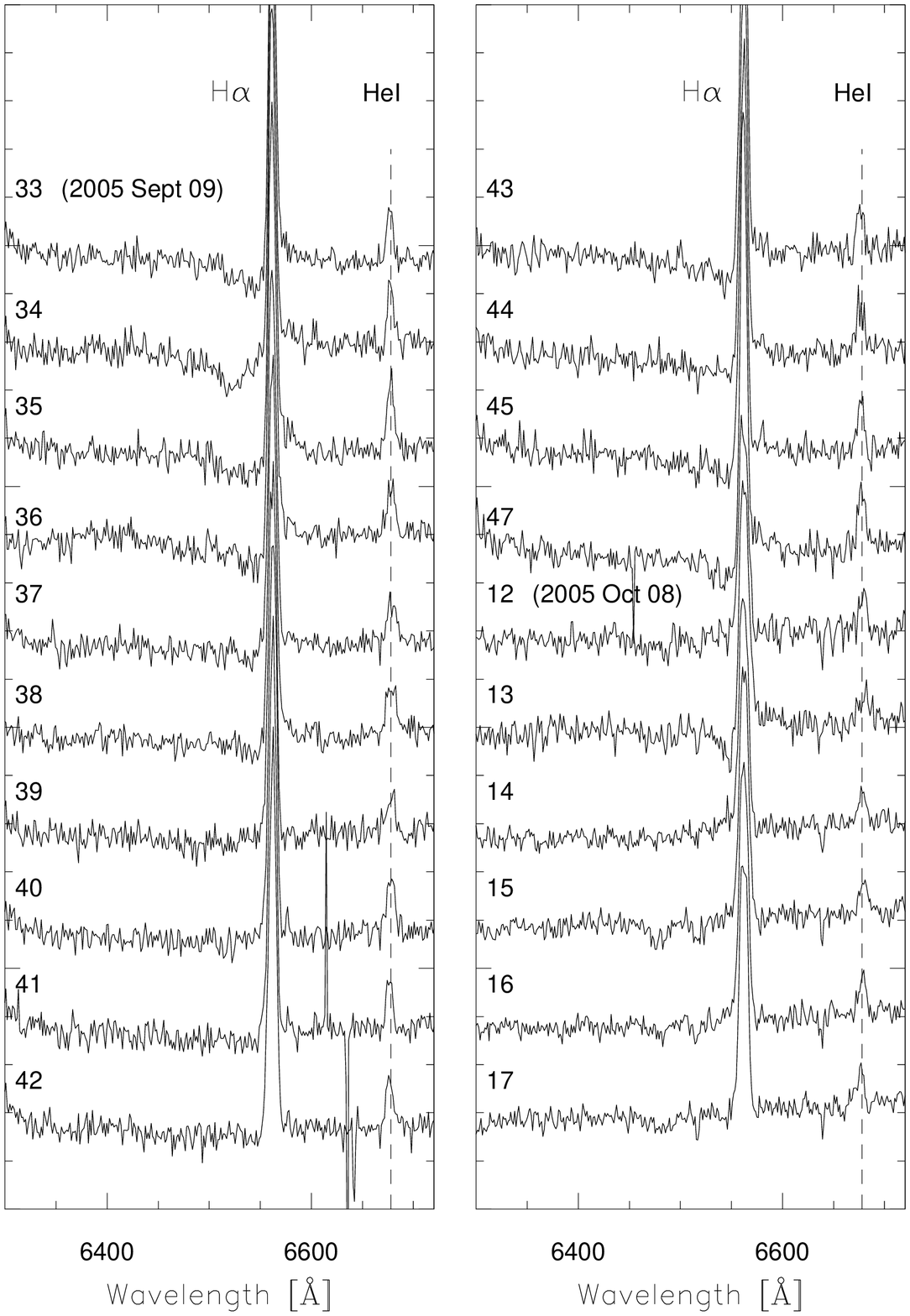}{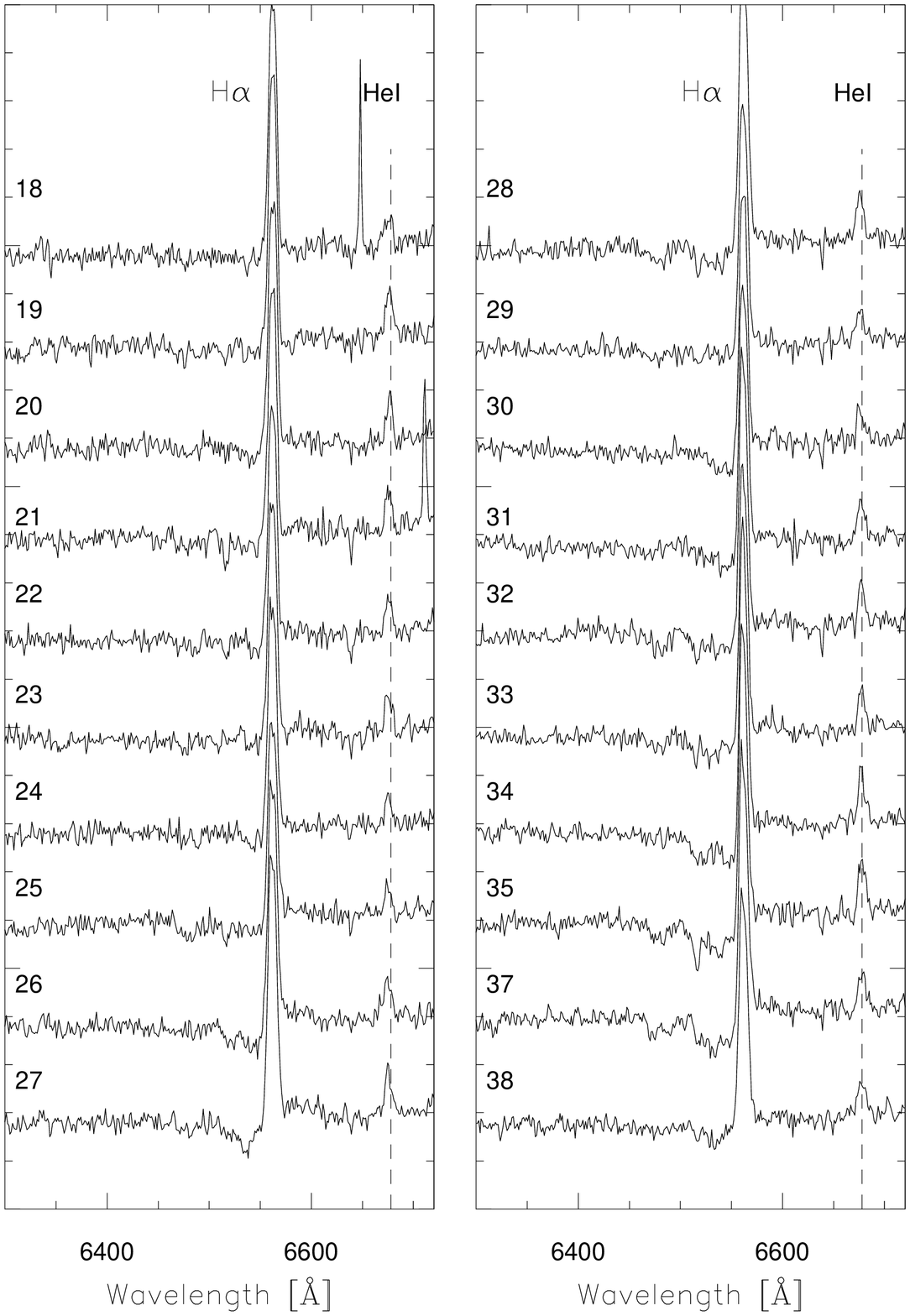}
\caption{Nested spectra around the H$\alpha$ line for all nights of our observations, demonstrating the time evolution of the H$\alpha$ P-Cygni profiles. The last spectrum from each sequence is omitted because of its low S/N\label{nested1}}
\end{figure}

Figure~\ref{nested1} shows the H$\alpha$ region in nested plots as a function of time for the 3 nights of our observations.  The typical spacing of the WIYN spectra in Figure~\ref{nested1} is 700 sec, while the typical spacing of the 2.1-m red spectra
    is 240 sec.  The spectral number is our internal numbering but is provided
    to denote the time order.  In these nested spectra we see that the blue
    absorption wing
    is not always present in H$\alpha$, consistent with the episodic character
    of the wind as seen in the optical (Kafka et al. 2003; Kafka \& Honeycutt
    2004; Kafka et al. 2004).  When present, the profile of the blue absorption
    wind changes, being sometimes narrow, sometimes broad, and occasionally
    having two components (for example, see spectra 25 and 33 in Figure~\ref{nested1}).
    The HeI 6678 emission line does not show P-Cygni, consistent with a
    disk origin of this line as deduced in Q Cyg (Kafka et al. 2003).
 No periodic or large amplitude EW variations are observed in any of the emission lines (see figures~\ref{HeI} and ~\ref{fullHa}); this is expected due to the low inclination of the system. An attempt to fold the RVs and/or EW of the lines to any of the system superhump periods lead to null results.

\subsection{An orbitally modulated outflow?}

 Figure 1 shows a P Cygni profile in the blue wing of 
the H$\alpha$ line; this is also seen in several of our 
individual spectra in Figures 5 and 6. 
Likewise, the \ion{He}{1} 5876 \AA\ emission line is 
occasionally accompanied by blueshifted absorption. 
Since the \ion{He}{1} 6678 \AA\ line does not display similar 
absorption, we can exclude the accretion disk and the WD 
as the origin of the absorption in \ion{He}{1}. 
The character of the wind in the \ion{He}{1} 5876 \AA\ seems 
to be different than in H$\alpha$; however, the presence of 
\ion{C}{4} 5801 \AA\ and 5812 \AA\ emission does not allow 
measuring wind velocity from the absorption blueward of \ion{He}{1} 5876\AA. 
In H$\alpha$ the wind shows multiple episodes reaching 
velocities of $Ã¢ÂÂ¼5000$ km s$^{-1}$. Because the blue absorption 
profile is both complicated and changeable, we simply integrated 
between $-2500$ and $1500$ km s$^{-1}$ to obtain 
an EW for the absorption; this is the same velocity range 
used in W03.  Figure 7 shows the results for our three nights of data.
The EW behavior is episodic, with relatively long intervals of no
measurable wind punctuated by a single active interval in each data set.
These activity episodes have 
typical durations of 30--40 min, similar to the H$\alpha$ wind 
event in Q Cyg (Kafka et al. 2003). In the W03 study, the blue 
absorption at H$\beta$ was found to be modulated with a period 
similar to the orbital period, with the difference tentatively 
attributed to wobbling of the accretion disk. In the FUSE far-UV 
spectra of V592 Cas, Prinja et al. (2004) found that the 
\ion{C}{3} 1176 $\AA$ line is strongly modulated with orbital 
phase, indicating a departure from axisymmetry in the outflow. 
They discuss possible explanations (e.g., a magnetic WD, 
overflowing gas in the disk, spiral density waves in the disk) 
without favoring any of them.

In the UV study of Prinja et al. (2004), the \ion{C}{3} 1176 $\AA$
line of V592 Cas displays its maximum EW for half the orbit. 
To investigate if the outflow in the optical is also 
phase-modulated, we binned and median combined our spectra in 
increments of 0.1 phase units.  We then looked for patterns with phase 
or time in the presence, variations, and strength of the absorption 
wind signature. Figure 7 shows that wind events during 
the three nights occurred near orbital phases 0.8, 0.9, and 0.1. 
Our individual data sets each span less than a complete orbital cycle, 
but the reoccurence of this behavior in three data sets obtained 
many orbital cycles apart suggests that it is a persistent 
orbital phase dependence.  Our spectra suggest that 
a similar phase-dependence is present in \ion{He}{1} 5876 \AA; however, the 
phase interval in which the outflow is present changes 
at different epochs. For example, all of the spectra 
from 2005 October 11 show the presence of an outflow 
associated with the \ion{He}{1} 5876 \AA\ line. 
This agrees with the behavior of the 
old nova Q Cyg, in which P Cygni profiles are always present in 
the \ion{He}{1} triplets, whereas they appear to be episodic in H$\alpha$ 
(Kafka et al. 2003). This also implies that the source 
of the outflow seen in these two lines is different. 
A possibility is that disk precession and/or spiral waves modulate 
the outflow axis (as suggested by W03); this would lead to 
periodic variations of the outflow on one of the superhump 
periods, which is not observed in our data (nor is demonstrated 
in the W03 data). The only sites in the system that can produce 
an outflow in which the wind propagation axis is tilted with 
respect to the orbital plane (thus ``breaking'' the wind axisymmetry) 
are the bright spots created by the initial impact of the stream on the 
accretion disk rim and/or at a secondary site within the disk in the case of
an overflowing stream (as suggested for the SW Sex stars; see Hoard et al. 2003 and references therein).
Both sites are energetically capable of producing an outflow 
with radial velocities visible only at certain orbital phases along 
the line-of-sight of the observer. 
Because we are unable to reliably associate 
orbital phase with viewing angle, we cannot discriminate 
between these two locations using our data. 

\subsection{Correlations with photometry?}

There is very little information on whether wind events in CVs 
are correlated with photometric variations, which could shed 
light on the origin of the wind. Kafka et al. (2003) suggested 
that enhanced wind events in Q Cyg might be correlated with 
stunted outbursts seen in V-band. Flickering time-scales in 
novalike CVs like V592 Cas overlap (at the slow end) with time 
scales of individual wind events. The origin of the flickering 
appears to be the inner disk in some CVs and the stream-impact 
bright spot on the disk in others (Welsh et al. 1997; Bruch 2000), 
locations that have also been suggested for the origin of the wind.  

During our 2005 Oct 8 spectroscopic observations of V592 Cas, we 
obtained simultaneous photometry covering $\sim 1.8$ hr using the 
WIYN 0.9m telescope. 
The average brightness of V592 Cas is $\langle V \rangle = 12.55$ 
with $\sim0.2$ mag quasiperiodic variations on time scales of minutes, 
similar to the behavior presented in Kato \& Starkey (2002). 
Periodogram analysis of our data and the white light AAVSO light 
curve of V592 Cas did not recover the 21-min period reported in 
Kato \& Starkey (2002) nor any other periodic signal. 
The left panels of Figure 9 show the individual spectra taken during our 
photometric observations.
Strong P Cygni profiles are present in all but one of the spectra ($\#$2), 
with some spectra ($\#$5, $\#$9) showing multiple P Cygni dips at different 
velocity offsets indicative of multiple successive outflow episodes. 
The spectra are numbered in sequence and these numbers are plotted at 
the times of observation in the top right panel of Figure 9, which 
also shows the V-band light curve.
We measured the maximum velocity of the absorption profiles in H$\alpha$
using the IRAF task splot and cursor positioning\footnote{With this 
method, the error in determining the maximum velocities in the 
spectra is of the order of 200 km s$^{-1}$, an order of 
magnitude smaller than the measured outflow velocities. 
Thus, the fact that these measurements 
were performed by visual inspection is of no consequence to the results.}. 
There is no consistently convincing correlation between the maximum 
velocity of the absorption and the brightness of the CV (middle right 
panel of Figure 9).
We also measured the EW of the wind absorption in the H$\alpha$ line 
using IRAF/splot (bottom right panel of Figure 9).
Again, there is no convincing correlation between the absorption EW and the
brightness of the system.  Taken together, these results imply that
that the characteristics of the wind (as expressed in H$\alpha$) are 
not related to the source of the V-band photometric modulations in V592 Cas.  

\section{Summary}

In this paper we present an optical spectroscopic study of V592 Cas, 
aimed at exploring the presence of an outflow in the optical spectra. 
In the red, the system has all the characteristics of a low 
inclination disk CV, with all the \ion{He}{1} and H$\alpha$ lines 
in emission. 
P Cygni profiles, indicative of an outflowing wind, are present in 
the blue wing of the H$\alpha$ 
and, more weakly, \ion{He}{1} 5876 \AA\ emission lines in many of our 
individual spectra; however no such absorption is present in the
\ion{He}{1} 6678 \AA\ line in any of our spectra.
Apparent variations in the systemic velocity of the 
emission line radial velocity curves likely stem from contamination 
of the emission line profiles by
a P Cygni component (i.e., wind) of different strength on different 
nights, rather than an actual change in the overall systemic velocity 
or accretion flow in the CV.
The wind in the H$\alpha$ line is episodic, with events having duration 
significantly less than the orbital period of the CV.  The observed 
maximum outflow velocities approach $\sim$5000 km s$^{-1}$.
Some of the P Cygni profiles show multiple dips at different velocities 
indicative of successive outflow events. There is some indication that 
the H$\alpha$ outflow is present only in a specific orbital phase range 
($\sim0.8$--$0.1$), in agreement with relevant studies in the UV 
suggesting that the wind in V592 Cas is phase-modulated. If the phase 
dependence of the P Cygni profiles is confirmed, then a possible origin 
for this component of the outflow is a bright spot on the edge and/or 
interior of the accretion disk. An observed phase shift between the 
RVs of \ion{He}{1} and the central H$\alpha$ line is similar to the 
characteristic behavior of the SW Sextantis stars (Hoard et al. 2003 
and references therein).  Definitive classification of V592 Cas as an 
SW Sex star is beyond the scope of our current data.  However, we note that 
this would be consistent with a possible scenario for the origin of 
the outflow in V592 Cas in a secondary bright spot in the disk interior 
at the impact site of an overflowing accretion stream.  
Simultaneous optical photometry during one night of our spectroscopic 
observations does not show any convincing correlation between the 
observed photometric variations and the presence or strength of the 
wind in V592 Cas. 

One of the aims of our work is to examine if the observed outflow can be accounted for angular momentum loss in this system. In the literature the relevant models calculating the mass and angular momentum loss rate due to outflows from CVs are applicable for species such as the UV SiV and CIV and take into account their ionization fraction, line shape and observed velocity in the magnetohydrodynamic computations. Those calculations use the P-Cygni absorption profiles in UV resonance lines, the mass of the white dwarf (usually assumed to be M$_{WD}$=0.6M$_{sun}$) a certain mass-radius relationship for the donor star and two free parameters, namely the $\dot{M}$ and the inclination {\it i} of the system, demonstrating the effects of different combinations of the latter two on the mass loss rate and angular momentum loss due to the wind (e.g., Proga \& Kallman 2002). Of the key assumptions of the calculations is that the wind is bipolar and axisymmetric, originating on the inner or outer part of the accretion disk. As suggested in this paper (and in Prinja et al. 2004), the wind in V592 Cas is phase modulated, indicating a source with a varying visibility. Thus the existing theoretical calculations for angular momentum loss in CVs can not be applied in V592 Cas. Therefore in V592 Cas we can not derive wind properties (such as the mass loss rate) and assess if those episodes can be responsible for angular momentum loss from the system. Since V592~Cas is not the only CV with phase-dependent outflows in the UV resonance lines (Prinja et al. 2004), there is a pressing need for a model that takes into account a non-axisymmetric wind in order to examine if such a wind can be responsible for mass and angular momentum loss in these systems driving their evolution.

\acknowledgments
We would like to thank our anonymous referee for her/his thorough review of the manuscript.

Facilities: \facility{KPNO 2.1m}, \facility{WIYN 3.5m}, \facility{WIYN 0.9m}.

\clearpage

\begin{figure}
\epsscale{1.3}
\plottwo{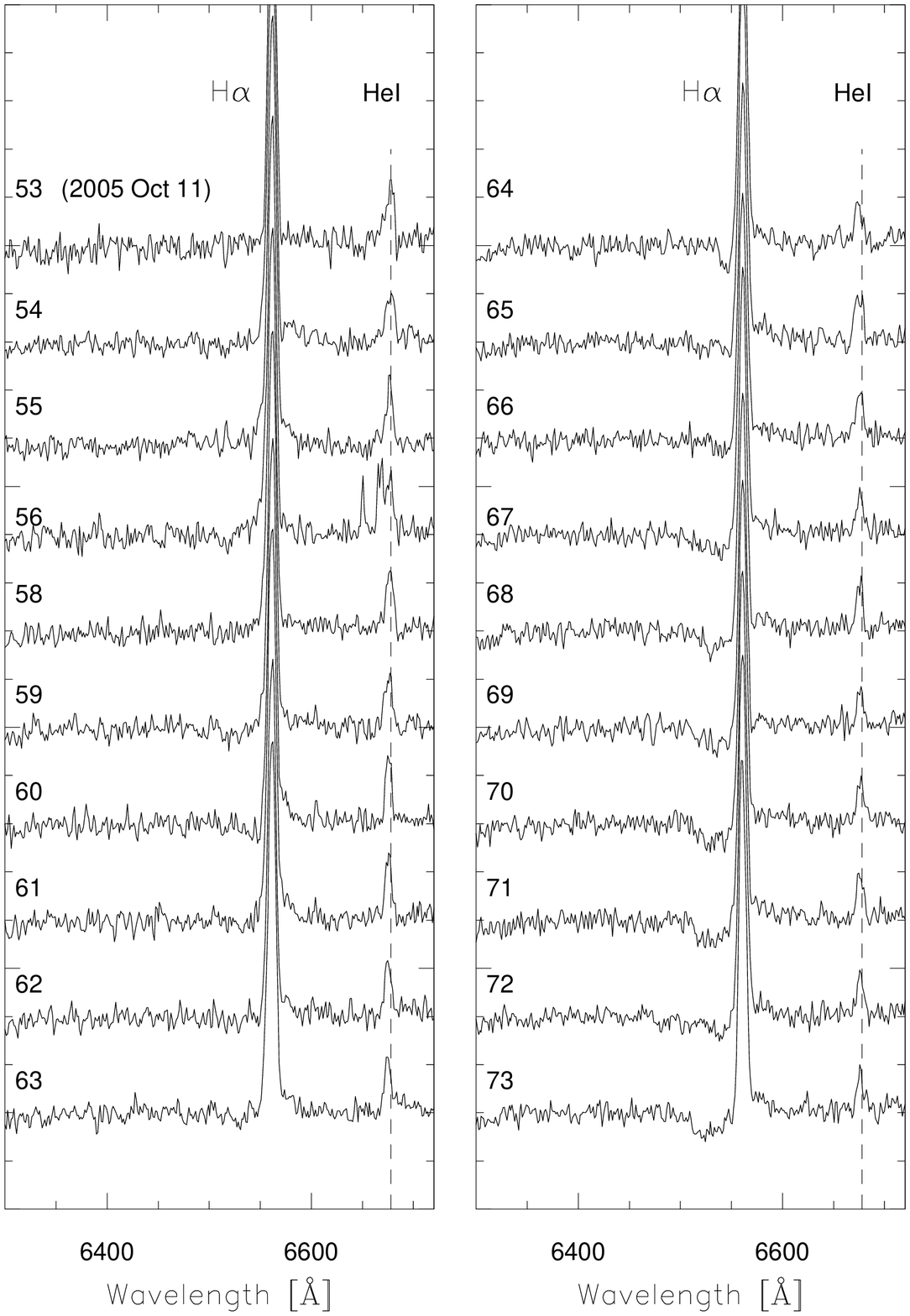}{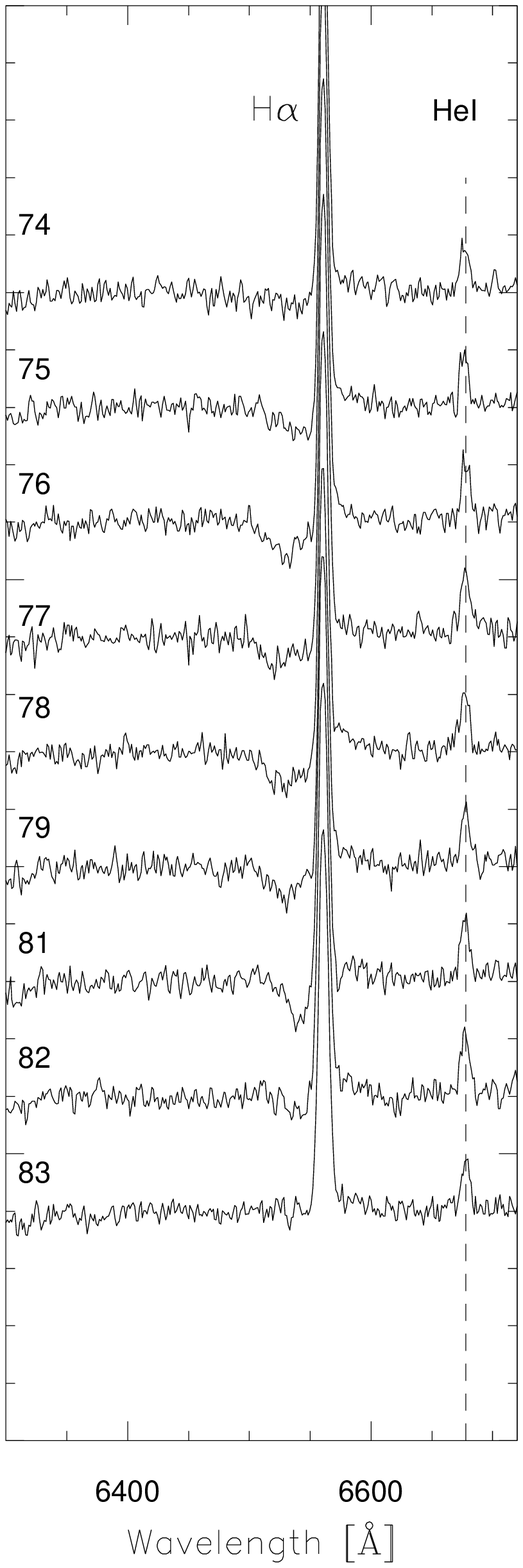}
\caption{Figure~\ref{nested1} - CONT.\label{nested2}}
\end{figure}

\begin{figure}
\epsscale{0.9}
\plotone{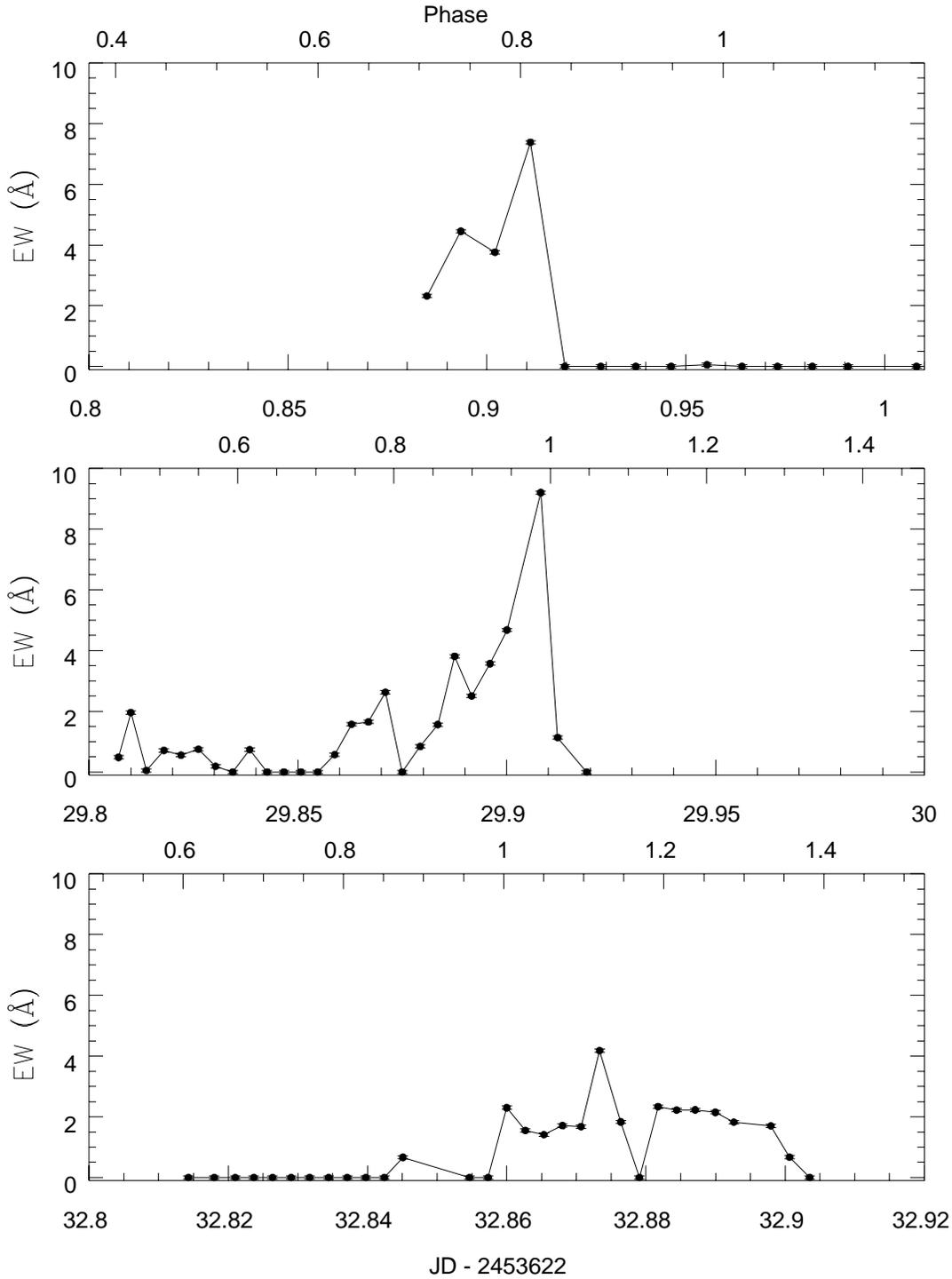}
\caption{Equivalent widths of the blueshifted absorption (P-Cygni) in the H$\alpha$ line for the three nights of our spectroscopic observations. The top axis in each panel corresponds to phase, whereas the bottom panel is time (JD-2453622). See text for discussion.\label{ewall}}
\end{figure}

\begin{figure}
\epsscale{1.2}
\plottwo{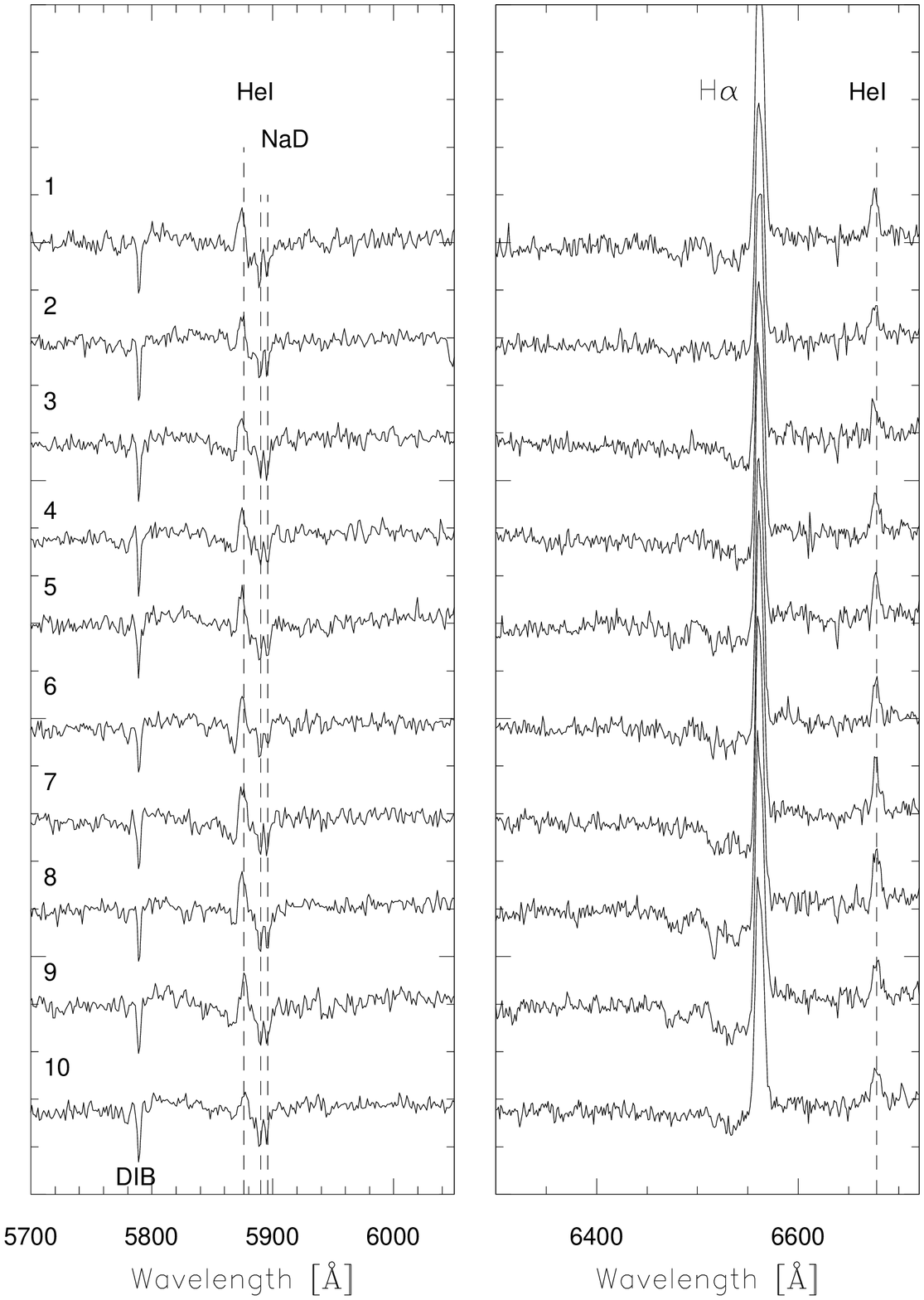}{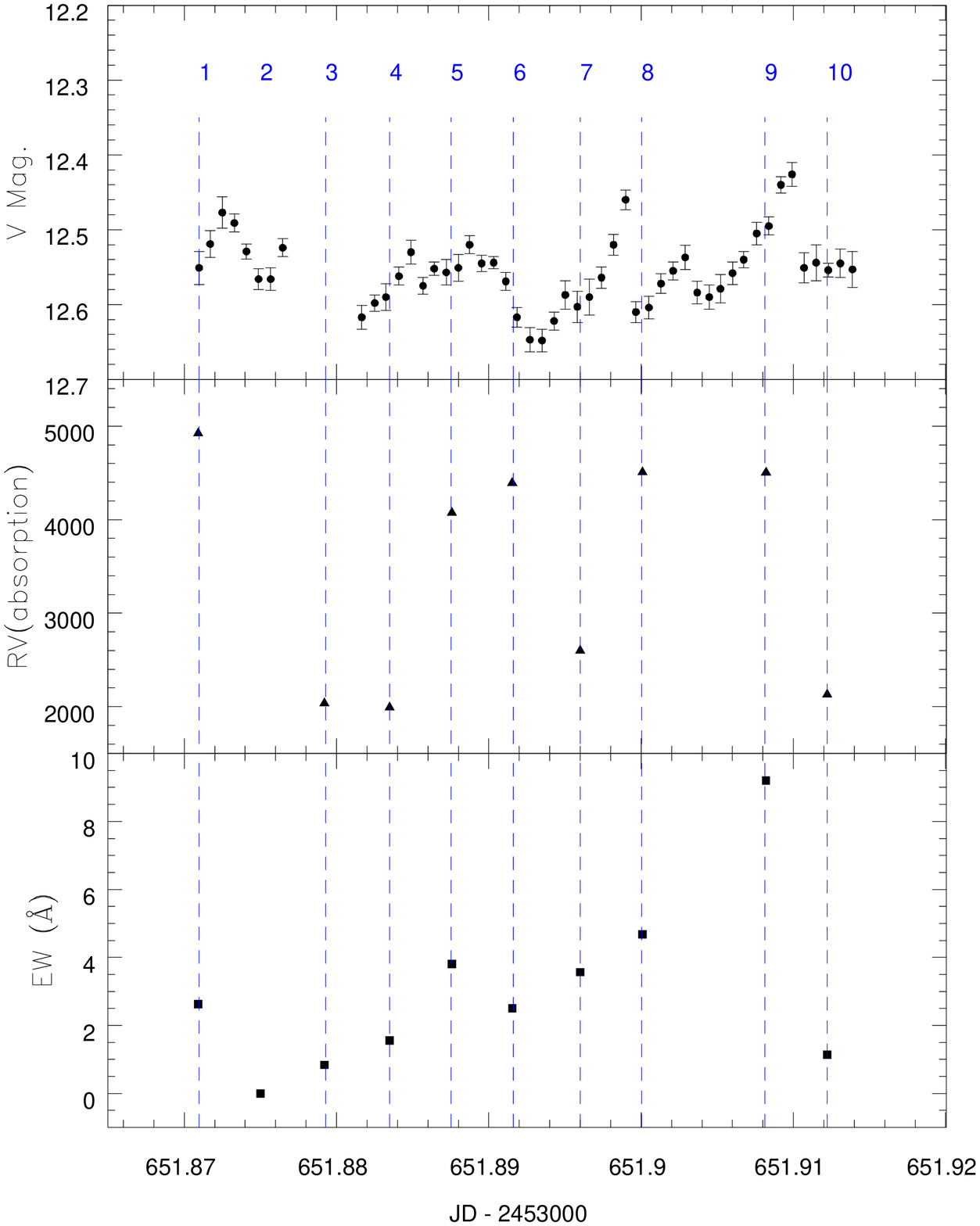}
\caption{
(left and middle) Spectra of V592 Cas at the time 
of the photometric observations. The spectra are corrected for telluric 
features, normalized to unity, and offset by a constant for clarity. 
(right, top) V-band light curve of V592 Cas.  The number labels 
correspond to the times of observation of the correspondingly numbered 
spectra in the left panels.
(right, middle) Maximum radial velocity of the blueshifted absorption 
(P Cygni) profiles at H$\alpha$.  
(right, bottom) Total EWs of the blueshifted absorption at H$\alpha$.  
The vertical dashed lines in the right panels show the times of 
observation of the spectra that display blueshifted absorption features 
at the H$\alpha$ and \ion{He}{1} 5876 \AA\ lines.\label{wind}}
\end{figure}

\clearpage
\begin{deluxetable}{ccccc}
\tablecolumns{5}
\tablewidth{0pc}
\tablecaption{Log of Observations of V592 Cas.\label{tbl-1}}
\tablehead{\colhead{UT Date} & \colhead{Telescope/} & \colhead{} & \colhead{No.} & \colhead{integration time} \\
 & instrument &  & exposures & (sec)} 
\startdata
Spectroscopy  &      & wavelength coverage &  &  \\
\hline
2005 Sept 09 & WIYN 3.5m/Hydra MOS & 5500-8000$\AA$ & 14 & 600 \\
2005 Oct 08  & KPNO 2.1m/GoldCam   & 5500-8000$\AA$ & 2 & 180 \\ 
            &                      & 5500-8000$\AA$ & 25 & 300 \\
2005 Oct 11 &                      & 5500-8000$\AA$ & 30 & 180 \\
\hline
\hline
Photometry   &      & Filter &  &  \\
\hline
2005 Oct 08   &  WIYN 0.9m/S2KB   & V & 58  & 30 \\
\hline
\enddata
\end{deluxetable}

\begin{deluxetable}{cccc}
\tablecolumns{4}
\tablewidth{0pc}
\tablecaption{Measurements of the emission lines in the average spectrum of V592 Cas.\label{tbl-2}}
\tablehead{\colhead{line} & \colhead{rest wavelength} & \colhead{gfwhm\tablenotemark{a}} & \colhead{EW\tablenotemark{b}}  \\
 & ($\AA$) &  & ($\AA$) }
\startdata
He I  &  5876   & 5.223  &  -0.354 \\
H$\alpha$  &  6563  &  8.790 & -6.078 \\
He I  &  6675  &   8.155 & -0.722 \\
He I  &  7065  &  6.926 & -0.480 \\
\enddata
\tablenotetext{a}{gfwhm=full width half maximum of the Gaussian fit}
\tablenotetext{b}{We follow the IRAF convention, where negative EW values correspond to the strength of emission lines.}
\end{deluxetable}

\begin{deluxetable}{cccc}
\tablecolumns{4}
\tablewidth{0pc}
\tablecaption{Fits on the RVs of the emission lines for the three epochs of observations}
\tablehead{\colhead{Line} & \colhead{$\gamma$(km/sec)} & \colhead{K(km/sec)} & \colhead{$\phi_{0}$}}
\startdata
H$\alpha$(center) &         &           &                  \\
2005 Sept 9     & 37$\pm$1 & 27$\pm$2  &  0.020$\pm$0.011  \\
2005 Oct 8      & 56$\pm$1 & 21$\pm$4  &  0.027$\pm$0.025  \\ 
2005 Oct 11     & 73$\pm$1 & 25$\pm$2  &  0.015$\pm$0.014  \\
\hline
HeI 5876$\AA$     &                 &                &       \\
2005 Sept 9     &   -64$\pm$2  & 54$\pm$11 & 0.237$\pm$0.026  \\
2005 Oct 8      &   -62$\pm$1  & 66$\pm$8  & 0.178$\pm$0.017  \\
2005 Oct 11     &   -124$\pm$2 & 57$\pm$7  & 0.202$\pm$0.014  \\
\hline
HeI 6678$\AA$     &                 &                &    \\
2005 Sept 9     &  -29$\pm$1 & 58$\pm$7 &  0.169$\pm$0.017 \\
2005 Oct 8      &  -51$\pm$1 & 78$\pm$7 &  0.173$\pm$0.012 \\
2005 Oct 11     &  -73$\pm$2 & 61$\pm$7 &  0.159$\pm$0.015 \\
\hline
\enddata
\end{deluxetable}

\clearpage
\begin{deluxetable}{ccccc}
\tablecolumns{5}
\tablewidth{0pc}
\tablecaption{Log of Observations of V592 Cas.\label{tbl-1}}
\tablehead{\colhead{UT Date} & \colhead{Telescope/} & \colhead{} & \colhead{No.} & \colhead{integration time} \\
 & instrument &  & exposures & (sec)} 
\startdata
Spectroscopy  &      & wavelength coverage &  &  \\
\hline
2005 Sept 09 & WIYN 3.5m/Hydra MOS & 5500-8000$\AA$ & 14 & 600 \\
2005 Oct 08  & KPNO 2.1m/GoldCam   & 5500-8000$\AA$ & 2 & 180 \\ 
            &                      & 5500-8000$\AA$ & 25 & 300 \\
2005 Oct 11 &                      & 5500-8000$\AA$ & 30 & 180 \\
\hline
\hline
Photometry   &      & Filter &  &  \\
\hline
2005 Oct 08   &  WIYN 0.9m/S2KB   & V & 58  & 30 \\
\hline
\enddata
\end{deluxetable}

\begin{deluxetable}{cccc}
\tablecolumns{5}
\tablewidth{0pc}
\tablecaption{Radial velocity measurements for our spectra}
\tablehead{\colhead{Spectrum ID\tablenotemark{a}} & \colhead{HJD-2453622} & \colhead{Phase} & \colhead{radial velocity}  \\
 & &  & (km/sec) }
\startdata
33  &  0.88  &  0.112  &  49.4  \\
34  &  0.89  &  0.186  &  61.0  \\
35  &  0.90  &  0.261  &  63.2  \\
36  &  0.91  &  0.338  &  63.2  \\
37  &  0.92  &  0.413  &  62.8  \\
38  &  0.93  &  0.492  &  32.0  \\
39  &  0.94  &  0.569  &  25.1  \\
40  &  0.95  &  0.646  &  8.3  \\
41  &  0.96  &  0.724  &  12.8  \\
42  &  0.96  &  0.800  &  11.5  \\
43  &  0.97  &  0.878  &  18.3  \\
44  &  0.98  &  0.954  &  27.3  \\
45  &  0.99  &  0.032  &  28.6  \\
47  &  1.01  &  0.182  &  53.5  \\
12  &  29.81  &  0.471  &  70.6  \\
13  &  29.81  &  0.497  &  30.2  \\
14  &  29.81  &  0.529  &  50.4  \\
15  &  29.82  &  0.565  &  20.9  \\
16  &  29.82  &  0.601  &  43.2  \\
17  &  29.83  &  0.637  &  34.8  \\
18  &  29.83  &  0.673  &  36.1  \\
19  &  29.83  &  0.709  &  26.8  \\
20  &  29.84  &  0.744  &  14.1  \\
21  &  29.84  &  0.780  &  28.1  \\
22  &  29.85  &  0.815  &  46.6  \\
23  &  29.85  &  0.851  &  57.9  \\
24  &  29.85  &  0.886  &  60.3  \\
25  &  29.86  &  0.921  &  54.4  \\
26  &  29.86  &  0.956  &  36.8  \\
27  &  29.87  &  0.991  &  56.6  \\
28  &  29.87  &  0.026  &  59.4  \\
29  &  29.87  &  0.061  &  69.9  \\
30  &  29.88  &  0.099  &  51.7  \\
31  &  29.88  &  0.135  &  56.5  \\
32  &  29.89  &  0.171  &  73.1  \\
33  &  29.89  &  0.206  &  69.3  \\
34  &  29.90  &  0.244  &  67.1  \\
35  &  29.90  &  0.279  &  71.8  \\
37  &  29.91  &  0.350  &  86.0  \\
38  &  29.92  &  0.385  &  78.6  \\
53  &  32.81  &  0.606  &  48.2  \\
54  &  32.82  &  0.639  &  52.9  \\
55  &  32.82  &  0.665  &  46.7  \\
56  &  32.82  &  0.688  &  48.4  \\
57  &  32.83  &  0.712  &  47.2  \\
58  &  32.83  &  0.735  &  45.8  \\
59  &  32.83  &  0.758  &  37.4  \\
60  &  32.83  &  0.781  &  39.8  \\
61  &  32.84  &  0.805  &  59.2  \\
62  &  32.84  &  0.828  &  49.1  \\
63  &  32.84  &  0.851  &  57.6  \\
64  &  32.85  &  0.874  &  60.3  \\
65  &  32.85  &  0.957  &  85.2  \\
66  &  32.86  &  0.981  &  64.4  \\
67  &  32.86  &  0.004  &  84.3  \\
68  &  32.86  &  0.027  &  72.6  \\
69  &  32.87  &  0.050  &  80.6  \\
70  &  32.87  &  0.073  &  79.6  \\
71  &  32.87  &  0.097  &  89.4  \\
72  &  32.87  &  0.120  &  87.8  \\
73  &  32.88  &  0.146  &  110.7  \\
74  &  32.88  &  0.170  &  104.4  \\
75  &  32.88  &  0.193  &  90.1  \\
76  &  32.88  &  0.216  &  96.5  \\
77  &  32.89  &  0.239  &  89.9  \\
78  &  32.89  &  0.264  &  91.7  \\
79  &  32.89  &  0.287  &  103.8  \\
81  &  32.90  &  0.334  &  107.5  \\
82  &  32.90  &  0.357  &  102.8  \\
83  &  32.90  &  0.382  &  92.8  \\
\enddata
\tablenotetext{a}{The spectrum ID is the same as the number presented in figure 5}
\end{deluxetable}

\end{document}